\documentclass[
    ,final            
  ]
  {aipproc}

\layoutstyle{6x9}

\begin{document}

\title{Strangeness on the Nucleon}


\author{R.~Gonz\'alez-Jim\'enez}{
  address={Departamento de F\'isica At\'omica, Molecular y Nuclear, 
  Universidad de Sevilla, 41080 Sevilla, Spain}
}

\author{J.A.~Caballero}{
  address={Departamento de F\'isica At\'omica, Molecular y Nuclear, 
  Universidad de Sevilla, 41080 Sevilla, Spain}
}

%
%
\maketitle



The last decade a great effort has been made both experimental and theoretical 
on the contribution of the strange quark-antiquark pair to the electroweak nucleon 
structure. 
The experiments on Parity Violation Elastic Electron-Proton Scattering (PVep): 
SAMPLE~\cite{SAMPLE}, HAPPEX~\cite{HAPPEX1}, 
PVA4~\cite{PVA41} and G0~\cite{G01}, 
shed light on the electroweak vector current by 
measuring the PV asymmetry. On the other hand, the quasielastic 
neutrino-nucleus scattering experiments (MiniBooNE~\cite{MiniBooNE}) allow us 
to explore the axial contribution to the Weak Neutral Current (NC).

\begin{figure}
  \includegraphics[height=.32\textheight,angle=270]{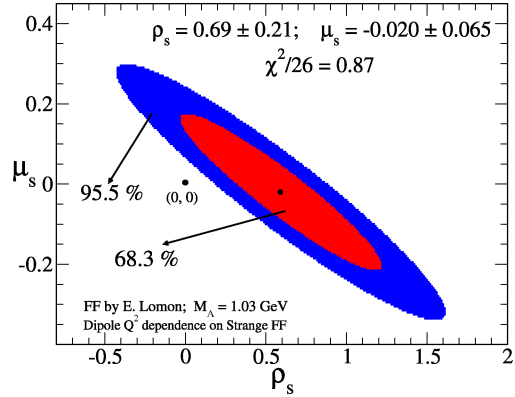}\label{mus_rhos}
  \caption{World data constraint in the $\mu_s-\rho_s$ plane. 
  1$\sigma$ (red) and 2$\sigma$ (blue) allowed region. See Ref.~\cite{GCD} 
  for details.}
\end{figure}

In Ref.~\cite{GCD} we review the state-of-art of the theoretical ingredients entering 
in the PV asymmetry in order to establish the confidence level contours for the
strange electric and magnetic parameters 
from a global analysis of the PVep asymmetry experimental data.
The specific strangeness content is
given by the static strangeness parameters $\rho_s$ and $\mu_s$ in the electric 
and magnetic sectors respectively.
The result of our $\chi^2$-analysis, for a specific set of 
theoretical inputs, 
is shown in Fig.~\ref{mus_rhos}. The ellipses (red and blue) represent the 
confidence contours around the point of maximum likelihood (black). 
The value of the point of maximum likelihood and the value of $\chi^2$ divided by the 
number of degrees of freedom for the system are also explicitely shown in the figure.
Although further studies and
investigations are needed before definite conclusions on the strangeness content 
in the nucleon can be drawn, the analysis of the 1$\sigma$ and 2$\sigma$ confidence 
ellipses shows that the case of
no strangeness, represented as (0,0) in this figure, 
is excluded by most of the fits (see \cite{GCD}).

\begin{figure}
  \includegraphics[height=.24\textheight,angle=0]{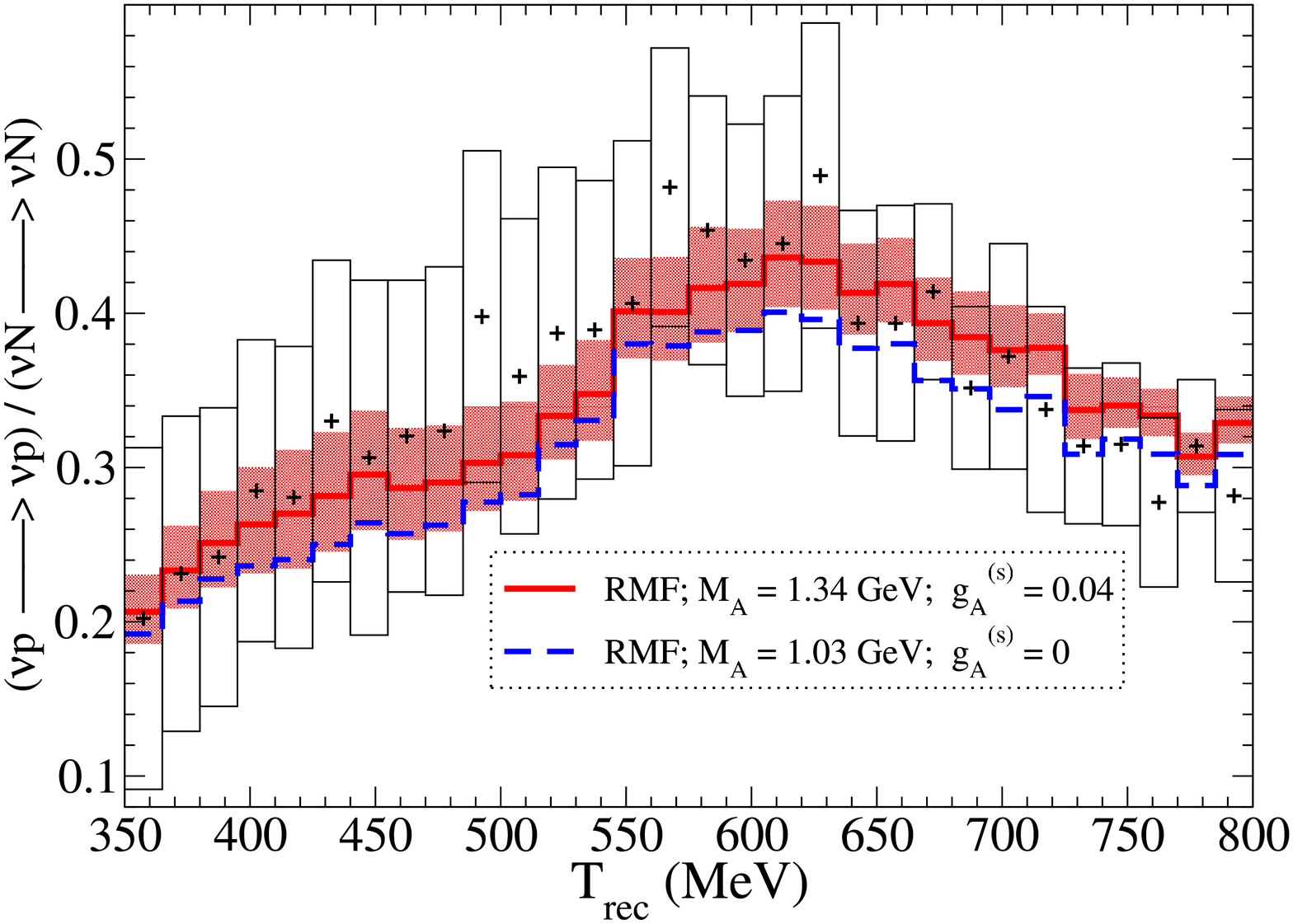}\label{ratio}
  \caption{The MiniBooNE data for the \textit{ratio}~\cite{MiniBooNE} 
          (black rectangles) are compared with our prediction within the RMF model. 
          We represent the 1$\sigma$ allowed region (red area) for the static 
          strange axial parameter, $g_A^{(s)}$, as well as the situation of 
          zero strangeness (blue line) as reference. See Ref.~\cite{GIBCU} for
          details.}
\end{figure}

In Ref.~\cite{GIBCU} the MiniBooNE NC data~\cite{MiniBooNE} are used to test the 
validity of the Relativistic Mean Field (RMF) and Superscaling Approach model 
(SuSA) in such experimental scattering situation. 
The NC quasielastic neutrino-nucleus cross section is largely affected 
by the axial contribution. However, it shows a very mild dependence on the 
axial strangeness. 
Thus, in Ref.~\cite{GIBCU}, the cross section data from the MiniBooNE experiment 
were used to
improve the description of the axial form factor by fitting the axial mass parameter.
Having controlled the axial form factor we studied the ratio of 
proton to neutron cross section. 
This observable presents a strong dependence on the strange axial form factor, 
while being almost independent on the specific model considered.
In Fig.\ref{ratio} we present the comparation between the experimental \textit{ratio} 
and our prediction. 
The values of the axial mass, $M_A$, and strange axial parameter, 
$g_A^{(s)}$, presented in this figure
are the result of a $\chi^2$-analysis in which the cross section 
and \textit{ratio} experimental data of Ref.~\cite{MiniBooNE} were used.

This work is supported by Spanish DGI (FIS2011-28738-C02-01). R.G.J. acknowledges 
support from Ministerio de Educaci\'on.
%


\bibliographystyle{aipproc}   

\begin{thebibliography}{9}

\bibitem{SAMPLE}
D. T. Spayde et al. [SAMPLE Collaboration], \emph{Phys. Rev. Lett.} \textbf{84}, 
1106 (2000).

\bibitem{HAPPEX1}
K.~A.~Aniol~et~al. [HAPPEX-99], \emph{Phys.~Rev.~C} \textbf{69}, 
065501 (2004); 
K.~A.~Aniol~et~al. [HAPPEX-a], \emph{Phys.~Lett.~B} \textbf{635}, 
275 (2006); 
A.~Acha~et~al. [HAPPEX-b], \emph{Phys.~Rev.~Lett.} \textbf{98}, 
032301 (2007); 
Z.~Ahmed~et~al. [HAPPEX-III], \emph{arXiv:1107.0913v1} 
[nucl-ex].

%
%

\bibitem{PVA41} 
F.~E.~Maas~et~al., [PVA4], \emph{Phys.~Rev.~Lett.} \textbf{93} 
(2004) 022002; 
F.~E.~Maas~et~al., [PVA4], \emph{Phys.~Rev.~Lett.} \textbf{94} 
(2005) 152001.


\bibitem{G01}
D. S. Armstrong et al. [G0 Collaboration], \emph{Phys. Rev. Lett.} \textbf{95}, 
092001 (2005); 
D.~Androi\'c~et~al. [G0 Collaboration], \emph{Phys. Rev. Lett.} \textbf{104}, 
012001 (2010).


\bibitem{MiniBooNE}
A. A. Aguilar-Arevalo et al. [MiniBooNE Collaboration], \emph{Phys. Rev. D}
\textbf{82} (2010) 092005.

\bibitem{GCD}
R. Gonz\'alez-Jim\'enez et al. 
\emph{Physics Reports} \textbf{524} (2013) 1.  
doi:10.1016/j.physrep.2012.10.003

\bibitem{GIBCU}
R. Gonz\'alez-Jim\'enez et al.,
\emph{Physics Letters B} \textbf{718} (2013) 1471.
doi:10.1016/j.physletb.2012.11.065

\end{thebibliography}

\end{document}